\documentclass[12pt,preprint]{aastex}
\received{}
\revised{}
\accepted{}

\shorttitle{High-z Quasars}
\shortauthors{Fang et al.}

\begin{document}

\title{{\sl Chandra} Observations of Two High-Redshift Quasars}

\author{Taotao Fang, Herman L.Marshall, Greg L. Bryan\altaffilmark{1} AND Claude R.Canizares}
\affil{Department of Physics and Center for Space Research}
\affil{Massachusetts Institute of Technology}
\affil{NE80-6081, 77 Massachusetts Avenue, Cambridge, MA 02139}
\altaffiltext{1}{Hubble Fellow}
\email{fangt@space.mit.edu, hermanm@space.mit.edu, gbryan@alcrux.mit.edu, crc@space.mit.edu}
	
\begin{abstract}

We report the first high-resolution X-ray spectra of two high-redshift quasars, S5 0836+710 and PKS 2149-306, obtained with the {\sl Chandra} High Energy Transmission Grating Spectrometer (HETGS). The primary goal of this observation is to use the high spectral resolving power of the HETGS to detect X-ray absorption produced by a hot intergalactic medium. The continuum of both quasars can be fitted by absorbed power laws. The power law photon index ($\Gamma$) of S5 0836+710 is consistent with that found in previous observations with {\sl ROSAT} and {\sl ASCA}, while the power law of PKS 2149-306 is harder than found in a previous observation by {\sl ASCA}. Excess continuum absorption above the Galactic value is found in S5 0836+710, as evidenced in {\sl ASCA} and {\sl ROSAT} observations. No significant emission or absorption feature is detected in either source at $\pm3\sigma$ level. Based on the detection limits we constrain the properties of possible emitters and absorbers. The upper limit of the equivalent width of Fe K emission lines could be as low as $\sim 10\ eV$. Absorbers with a column density higher than $8\times10^{16}\ cm^{-2}$ for \ion{O}{8} or $5\times10^{16}\ cm^{-2}$ for \ion{Si}{14} would have been detected. We propose a method to constrain the cosmological parameters (namely $\Omega_{0}$ and $\Omega_{b}$) via the X-ray forest theory, but current data do not give significant constraints. However, we estimate that observing $\sim 7$ bright quasars should give at least one \ion{O}{8} and more \ion{O}{7} absorption lines at $95\%$ possibility. We also find that combined with the constraints from the distortion of the CMB spectrum, the X-ray Gunn-Peterson test can marginally constrain a uniform, enriched IGM.

\end{abstract}

\keywords{intergalactic medium --- quasars: absorption lines --- quasars: individual (S5 0836+710, PKS 2149-306) --- cosmology: observations --- X-rays: galaxies}

\clearpage

\section{Introduction}

Cosmological numerical simulations suggest that a large fraction of the baryonic matter should reside in intergalactic space in the form of a hot/warm intergalactic medium (IGM) (see, e.g., \citealp{fhp98, cos99, dco00}). This hot/warm IGM contains at least $30\%-40\%$ of the baryons predicted by the standard theory of big bang nucleosynthesis (BBN) at temperatures between $10^{5}$ to $10^{7}$ K. Observationally little is known about this IGM. Recent observations of \ion{O}{6} absorption lines (see, e.g., \citealp{tsj00,tsa00,ots00}) indicate that these \ion{O}{6}-absorbers could be a significant ``baryon reservoir'' at low redshift.

Detecting the X-ray absorption features in the X-ray spectrum of a distant quasar was first proposed by \citet{sba80}. They suggested that an intervening uniform, hot IGM would introduce an ``absorption trough'' in the X-ray continuum spectrum of a distant quasar via the ``X-ray Gunn-Peterson test''. Absence of such an ``absorption trough'' would put constraints on the temperature, density and ionization states of the diffuse IGM. \citet{aem94} extended this idea by including photoionization from the X-ray background radiation. By using numerical simulations \citet{hgm98} explored the idea that a clumpy, highly-ionized IGM concentrated in the gravitational potential caused by dark matter might introduce resonant absorption lines in the X-ray spectrum of a background quasar and denoted them as the ``X-ray Forest'', in analogy to the Ly$\alpha$ forest in the optical/UV bands. Similar ideas were discussed by \citet{plo98} and \citet{fca00} (Hereafter FC).

A rough estimation shows that even for the largest mass concentrated objects, such as clusters of galaxies, the maximum column densities of the absorption ions (such as \ion{O}{8} and \ion{Si}{14}) are between $10^{16}$ - $10^{17}\ cm^{-2}$ and the maximum equivalent width (EW) is far below $10\ eV$ (see, e.g., \citealp{sar88,dav00}). All previous observations have been limited by the spectral resolving power. For instance, {\sl ROSAT} and {\sl ASCA} could not detect any spectral feature with an EW smaller than $10\ eV$. However, the launch of the {\sl Chandra} X-ray Observatory\footnote{See http://chandra.harvard.edu/} and {\sl XMM-Newton} make it possible to study the X-ray spectrum of a distant quasar with high-resolution spectroscopy. The High Energy Transmission Grating Spectrometer (HETGS, see \citealp{mcd94,can00}) on board {\sl Chandra} provides excellent spectral resolution with $\Delta E \approx 1\ eV$ around $1\ keV$. Given such a energy resolution, it is possible to study the resonant absorption lines or edges produced by the hot IGM in the X-ray spectrum of a distant quasar \citep{cfa98}.

In this paper we present observations of two high-redshift quasars: S5 0836+710 ($z=2.17$) and PKS 2149-306 ($z=2.34$). At redshift higher than $2$, these two sources are the brightest (S5 0836+710) and the second (PKS 2149-306) brightest known quasars in X-ray band\footnote{see the data archive under http://heasarc.gsfc.nasa.gov/}. Given such a high redshift, the light from the source traverses approximately $80\%$ of the Universe, which greatly increases the possibility of being attenuated by the intervening systems. Our primary goal is to find if the predicted ``X-ray forest'' is detectable with the {\sl Chandra} HETGS and if not, what are the limits of possible absorbers.

This paper is organized as follows: we describe the observations in \S\ref{sec:co} and the data reduction and analysis in \S\ref{sec:da}. In \S\ref{sec:xcim} we discuss the constraints on the IGM, the last section is a summary. Throughout the paper unless specifically mentioned, we use a Hubble constant of $H_{0} = 70h_{70}\ km\ s^{-1}Mpc^{-1}$.

\section{{\sl Chandra} Observation \label{sec:co}}

S5 0836+710 and PKS 2149-306 were observed with the High Energy Transmission Grating Spectrometer on board {\sl Chandra}. Detailed target information is listed in Table~\ref{t1}. The observation of PKS 2149-306 was broken into two parts, with approximately 18 hours between them. The HETGS produces a zero order image at the aim point of the focal plane detector, the ACIS-S array, with higher order spectra dispersed to either side (for ACIS-S, see \citealp{gar00}). The moderate energy resolution of the ACIS-S is used to separate the overlapping orders of the dispersed spectrum. The HETGS consists of two different grating assemblies, the High Energy Grating (HEG) and the Medium Energy Grating (MEG), and provides nearly constant spectral resolution ($\Delta\lambda = 0.012 \AA$ for HEG and $\Delta\lambda = 0.023 \AA$ for MEG) through the entire bandpass (HEG: $0.8$-$10\ keV$, MEG: $0.4$-$8\ keV$). 

\placetable{t1}

The ACIS-S array consists of 6 chips (S0-S5); however, due to a temporary malfunction of a portion of the on-board readout electronics (the FEP problem, see {\sl Chandra} Proposers' Observatory Guide, or POG), only five chips could be used (S1-S5). Due to the radiation-induced charge transfer inefficiency (CTI) the energy resolution of the four front-illuminated chips has become a function of the chips' row number, with a better resolution near the read-out nodes. Therefore, we moved the nominal aim point closer to the read-out nodes by about $3.3\ mm$ for S5 0836+710 and $2.4\ mm$ for PKS 2149-306 to get better ACIS energy resolution. The resulting ACIS resolution is adequate to allow unambiguous separation of the spectral orders.

Data were analyzed with the standard pipeline for the {\sl Chandra} HETGS provided by the Chandra X-ray Center (CXC)\footnote{See http://asc.harvard.edu/}. We use a combination of {\sl Chandra} Interactive Analysis of Observations (CIAO) V1.1 and custom routines in IDL. The standard screening criteria were applied to the data. We selected photon events with {\sl ASCA} grades 0, 2, 3, 4, 6 and excluded those with energies above $10\ keV$. We also removed hot columns and bad pixels in each CCD chip. Figure 1 displays the zeroth-order images of both targets in a $50^{\prime\prime} \times 50^{\prime\prime}$ region. We have smoothed the images in Figure 1 with a Gaussian ($\sigma=1.2^{\prime\prime}$) for visual clarity. Both unsmoothed zeroth-order images are consistent with point sources and can be well fitted by a single Gaussian with $\sigma \approx 0.38^{\prime\prime}$ for S5 0836+710 and $\sigma \approx 0.35^{\prime\prime}$ for PKS 2149-306. The light curves derived from the zeroth orders of both sources were constant to within statistics.

\placefigure{f1}

S5 0836+710 was reported as having a one-side, superluminal radio jet \citep{khq90,lkw98} at a position angle (PA) of $-146^{\circ}$. The jet has a length of at least $1.5^{\prime\prime}$, which is at the spatial resolving limit of the ACIS-S. The head of the jet is located between $1-2^{\prime\prime}$ from center with an opening angle of $\sim 2^{\circ}$. We examine this region in the zeroth order image and find that the $3\sigma$ upper limit of the flux from this region is $\lesssim 0.04\%$ of the total flux from the source. Compared with $0.4\%$ and $6\%$ in the jets of 3C273 \citep{mar00} and PKS 0637-752 \citep{sch00}, respectively, we conclude that no X-ray jet was found during this observation.   
The observation of PKS 2149-306 was separated into two parts and the zeroth order images show that the aim points of the two observations were shifted by approximately 2 pixels. We analyzed the two observations separately and summed them at level 2 of the CXC pipeline to get a combined spectrum.

\section{Data Analysis \label{sec:da}}

\subsection{Continuum Fitting}

To fit both MEG and HEG spectra, we ignored the $1-2\ keV$ portion of the MEG spectrum due to unmodelled systematic calibration errors of $\sim\ 10\%-15\%$ and fit the remaining part simultaneously with the HEG spectrum. We present the fitted HEG spectra in figure~\ref{f2}. The spectral model is a single absorbed power-law where the absorption cross section and metal abundances are from \citet{mmc83}. Table~\ref{t2} gives the best-fit parameters. All errors are quoted at $90\%$ confidence. For PKS 2149-306, since fits with $N_{H}$ as a free parameter do not converge, we fix $N_{H}$ at the Galactic value.

\placetable{t2}
\placefigure{f2}


S5 0836+710 was previously observed by both {\sl ROSAT} and {\sl ASCA} \citep{blw94,cmc97}. The {\sl ASCA} spectra gave a photon index ($\Gamma$) of $1.45^{+0.05}_{-0.05}$ or $1.32^{+0.03}_{-0.03}$ between $2$ and $10\ keV$, with free $N_{H}$ or $N_{H}$ fixed at the Galactic value, respectively. Two {\sl ROSAT} observations showed a steeper spectrum (the photon indices are $1.51^{+0.13}_{-0.12}$ and $1.55^{+0.22}_{-0.22}$ between $0.1$ and $2\ keV$, respectively). The photon index in this observation ($\Gamma = 1.388 \pm 0.012$) is consistent with that of {\sl ASCA} but is flatter than the {\sl ROSAT} result. We find excess continuum absorption (above the Galactic value, at $\Delta N_{H} \sim 4 \times 10^{20}\ cm^{-2}$). The Galactic hydrogen column density is $2.83 \times 10^{20}\ cm^{-2}$ with an $1\sigma$ error of $1 \times 10^{19}\ cm^{-2}$\citep{mll96}. Excess absorption was also found in {\sl ASCA} observation at $\Delta N_{H} \sim 8.5 \times 10^{20}\ cm^{-2}$ \citep{cmc97}. We see a decreasing of excess absorption of $\sim 4.5 \times 10^{20}\ cm^{-2}$ in a timescale of $\sim 1.5\ yr$ in quasar frame. It is still unclear the causes of the excess absorption (instrinsic to the quasar or due to the intervening systems) and its variation (see discussions in, e.g., \citealp{cmc97, rtu00}). The flux of S5 0836+710 decreased by a factor of approximately 2 between the two {\sl ROSAT} observations, and the {\sl ASCA} flux ($\sim 1.4\times 10^{-11}\ ergs\ cm^{-2}s^{-1}$ between 2 and 10 $keV$) is consistent with the lower {\sl ROSAT} flux. These observations show a flux approximately twice that seen with {\sl ASCA}, which is consistent with the higher {\sl ROSAT} flux. Given the overlap of the {\sl Chandra} HETG and {\sl ASCA} bandpass and the small instrumental calibration uncertainties of {\sl Chandra}, we conclude that the different flux levels are intrinsic to the quasar.

PKS 2149-306 was also observed by {\sl ROSAT} and {\sl ASCA} \citep{smb96,cmc97}. The flux measured using {\sl ASCA} between 2 and 10 $keV$ is $9.94 \times 10^{-11}\ ergs\ cm^{-2}s^{-1}$ and is about $30\%$ higher than our result. The photon indices measured using {\sl ASCA} ($1.55^{+0.07}_{-0.06}$ for detector SIS and $1.54^{+0.09}_{-0.09}$ for detector GIS) are significantly higher than our value ($\Gamma = 1.255 \pm 0.020$), which indicates spectral variability in this source. This source was also observed during the {\sl ROSAT} All-Sky Survey, but the small number of counts precluded spectral fitting; a rough comparison with {\sl ASCA} also implies spectral variability \citep{smb96}.

Optically selected quasars can be divided into two categories according to their radio loudness: radio-loud quasar (RLQ) or radio-quiet quasar (RQQ) (see, e.g., \citealp{ata86}). Both of the sources we observed are RLQs. Previous observations by {\sl Einstein}, {\sl ROSAT} and {\sl ASCA} showed that the two types of quasars have significantly different X-ray properties where typically RLQs have flatter power-law indices ($\Gamma \sim 1.6$) than RQQs ($\Gamma \sim 1.9$)(see, e.g., \citealp{rtu00}). Our observation are consistent with this finding, although the spectra we observed are significantly harder than the average.

\subsection{Line Analysis \label{sec:la}}

We are interested in any possible line features in the quasar spectra. To identify line features, we need a good measurement of the continuum. We use the above absorbed power law models for HEG spectra. For MEG spectra, we use the same models and fit the residuals with a 5-order polynomial to account for the remaining uncertainty in the calibration. This fitted continuum serves as a reference point for measuring line features. Since the MEG spectrum spans from $2\AA$ to $26\AA$, the 5-order polynomial will affect spectral features larger than $5\AA$ but will preserve narrow line features. The MEG spectra for both sources are shown in figure~\ref{f3}. The bottom panel of each plot gives $\chi$, and the two dotted lines within the $\chi$ plot correspond to $\pm3\sigma$ level where $\sigma$ is the square root of the observed counts in each bin.   

\placefigure{f3}

If the model fits the observation well, the counts in each bin should follow the Poisson statistic, with a mean count from the fitted spectrum. To verify this, we calculate the Poisson probability $P_{x}(\ge k)$ for each bin, where $x$ is the expect mean number and $k$ is the observed value. $P_{x}(\ge k)$ should uniformly distribute between $0$ and $1$, given that the model fits the observed spectrum. In figure~\ref{f4} we show the cumulative distributions of $P_{x}(\ge k)$ for both quasars. We also show the cumulative distribution of the uniform distribution as the dotted line. We run a Kolmogorov-Smirnov (KS) test to test the null hypothesis that the both $P_{x}(\ge k)$ are uniformly distributed. The KS test shows that the null hypothesis is accepted at a significance level of $0.59$ for S5 0836+710 and $0.32$ for PKS 2149-306.

\placefigure{f4}

A line feature should at least have a signal-to-noise ratio at or above $\pm3\sigma$ level to be identified. Figure~\ref{f3} shows that most of the bins are well distributed within the $\pm3\sigma$ level. The two exceptions occur at around $22.6 \AA$ for S5 0836+710, where $\chi=3.52$, and at around $17.76 \AA$ for PKS 2149-306, where $\chi=3.51$. Assuming a Poisson distribution for each bin, we ran $1,000$ Monte-Carlo simulations for both spectra and find that the probability for observing one bin with $\chi \ge 3.52$ in one observation of S5 0836+710 is $63.5\%$, and observing one bin with $\chi \ge 3.51$ in one observation of PKS 2149-306 is $48.9\%$. So we conclude that no line feature is identified.
	
\subsubsection{Emission Lines}

Iron emission lines have been detected in several high-redshift quasars with {\sl ASCA} (see \citealp{rtu00} and references therein). However, no previous observation discovered any emission lines in these two sources, with the exception of \citet{ygn99} (Y99 hereafter) for PKS 2149-306, which we will discuss later. In {\sl ASCA} observations the upper limit on the equivalent width of an Fe K$\alpha$ emission line ($\sim 6.4\ keV$) is $110\ eV$ for S5 0836+710 or $85\ eV$ for PKS 2149-306, respectively (\citealp{cmc97}; line measured in the observer's frame, with line width $\sigma = 0$ for an unresolved line). 

The Fe K emission lines are located between $6.4\ keV$ and $6.9\ keV$, ranging from neutral iron to helium or hydrogen-like iron. For both sources these lines are expected at energies around $2\ keV$ in the observer's frame, where the {\sl Chandra} HETGS has a large collecting area and high resolving power. No significant line feature was detected in the spectrum of either source at a $3\sigma$ level. In table~\ref{t3} we list the upper limit equivalent width (EW) at two energies, $6.4\ keV$ and $6.9\ keV$, for both MEG and HEG spectra for two assumed intrinsic line widths, $\sigma = 0$ and $\sigma = 0.1\ keV$, respectively. The instrumental line response function is taken into account. Since the {\sl Chandra} HETGS has better resolving power than {\sl ASCA}, we can put a much tighter constraint on the equivalent width of the iron emission lines. We should emphasize that the low value of $\sigma = 0$ is adopted to calculate the theoretically low limit of detectable EW and since we have no constraint on the upper limit of the intrinsic line width, the EW limit could be worse by a factor of 2 if $\sigma = 0.5\ keV$. 

\placetable{t3}

Y99 reported that they found a significant emission feature around $5\ keV$ in the {\sl ASCA} spectrum of PKS 2149-306. They proposed that this line was blue-shifted Fe-K emission with a bulk-motion velocity around $0.73c$ (head-on) and a rest-frame EW of $300\pm200\ eV$ (assuming an intrinsic line width of $0.01\ keV$, errors are $90\%$ confidence). We do not detect any emission feature around $5\ keV$ at $3\sigma$ level in either HEG or MEG spectra, and set an upper limit of the EW as low as $67\ eV$ (Table~\ref{t3}). If we adopt the value of $\sigma = 0.01\ keV$ as used by Y99, the rest-frame EW limit is $75\ eV$ and still much smaller than $100 eV$, which is their $90\%$ confidence lower bound. A recent observation by {\sl BeppoSAX} \citep{efs00} also addressed the existence of such a feature.
	
\subsubsection{Absorption Lines \label{sec:al}}

X-ray resonant absorption lines are produced by highly ionized heavy elements. The detection of any absorption line depends on three factors: (1) the spectral intensity of the background quasar; (2) the physical properties of the intervening absorbers, specifically, the ion column density and velocity dispersion (both turbulent and thermal broadening); (3) the instrumental properties, such as the resolving power and effective area. Given that (1) and (3) are already known, we can constrain the properties of the absorbers. For demonstration we select the the spectrum of S5 0836+710 and calculate the expected signal-to-noise ratio (S/N) for \ion{Si}{14} and \ion{O}{8} (see figure~\ref{f5}), given the column density of \ion{Si}{14} and \ion{O}{8} at different redshifts and velocity dispersions ($b$, in units of $km\ s^{-1}$). Two values of $b$-parameter, $200$ and $500\ km\ s^{-1}$, are adopted to bracket the typical velocity dispersions of hot gas associated with groups and clusters of galaxies. Since  the strongest resonance lines of \ion{Si}{14} and \ion{O}{8}, the Ly$\alpha$ transitions, are located at $2.01\ keV$ and $0.654\ keV$ (rest-frame), respectively, we calculate the S/N of \ion{Si}{14} based on HEG (HEG has a better energy resolution there) and the S/N of \ion{O}{8} based on MEG (HEG has no effective area below $0.8\ keV$). The maximum redshift of \ion{O}{8} is $0.3$ because MEG has negligible effective area below $0.5\ keV$. The horizontal line in figure~\ref{f5} is the $3\sigma$ detection level. The column density of \ion{O}{8} or \ion{Si}{14} must be at least $8 \times 10^{16}\ cm^{-2}$ or $5 \times 10^{16}\ cm^{-2}$, respectively, to produce an absorption line at or above the $3\sigma$ level. 

\placefigure{f5}

\section{X-ray Constraints on the Intergalactic Medium \label{sec:xcim}}

\subsection{X-ray Forest}


The detection of the X-ray resonant absorption lines is closely related to the overall baryon density ($\Omega_{b}$) and matter density ($\Omega_{0}$) in the Universe. For a given level of metal enrichment, higher baryon density in general implies larger amounts of highly ionized metals and a higher possibility of detecting the X-ray absorption lines created by these metals in the spectrum of a distant quasar. On the other hand, as we predicted in \citet{fbc00}, most of the high column density ions (specifically, at column density higher than $10^{15}\ cm^{-2}$) are located in the collapsed, virialized objects, such as groups or clusters of galaxies, in the form of hot intracluster or intragroup gas\footnote{As we will discuss in the next section, numerical simulations show that the X-ray absorption material may also arise from filaments connecting the virialized structures; however, the resonant absorption lines produced by the low-column density ions ($ < 10^{15}\ cm^{-2}$) from these filaments are generally too weak to be detectable with {\sl Chandra} and even {\sl Constellation-X}.}. Since $\Omega_{0}$ governs the overall amplitude of the density fluctuations, changing $\Omega_{0}$ will dramatically change the spatial density of virialized objects, and so the distribution of ions at high column density.

To investigate how $\Omega_{b}$ and $\Omega_{0}$ can affect the X-ray spectra of high redshift quasars, we calculate the X-ray forest distribution function (XFDF, for details, see \citealp{hgm98,plo98}; FC). The XFDF, $\partial^{2}P/\partial N^{i}\partial z$, is defined as the number of absorption systems along the line of sight with a column density between $N^{i}$ and $N^{i}+dN^{i}$ per unit redshift, where $N^{i} = N(X^{i})$ is the column density for ion $X^{i}$. For a source at redshift $z_{0}$, the total absorption line number for ion $X^{i}$, with a column density above a certain number $N^{i}_{0}$ is given by
\begin{equation}
\label{eq:exp}
n = \int_{0}^{z_{0}} dz \int_{N^{i}_{0}}^{\infty} dN^{i} \frac{\partial^{2}P}{\partial N^{i}\partial z}
\end{equation} Since the XFDF is determined by cosmological parameters, such as $\Omega_{0}$ and $\Omega_{b}$ (see figures 2,3,4 of FC), by comparing the expected absorption line number ($n$) with observations, we may set the constraints on these cosmological parameters. 

Based on the {\sl Chandra} observations we set the upper limit of different ion column densities (see section~\ref{sec:al}), which gives $N^{i}_{0}$ in equation~(\ref{eq:exp}). The minimum detectable column density for \ion{O}{8} is $N^{i}_{0} = 8 \times 10^{16}\ cm^{-2}$ and for \ion{Si}{14} is $N^{i}_{0} = 5 \times 10^{16}\ cm^{-2}$. We find, even in the extreme case, where $\Omega_{0}=1$ and $Z_{\odot} = 1$, the expected absorption line number for \ion{O}{8} absorption lines is only $n=0.65$ and even less for \ion{Si}{14}. Here $Z_{\odot}$ is the metal abundance relative to the solar value where $Z_{\odot}$ = 1 indicates solar abundance, and we integrate equation (\ref{eq:exp}) to $z_{0} = 0.3$ for \ion{O}{8} and to $z_{0} = 2.17$ for \ion{Si}{14}. The Poisson statistic shows that the probability of detecting less than one \ion{O}{8} absorption line is $0.53$ given such an expectation. So with the current observations we are not able to constrain those cosmological parameters. This is also evidenced in FC. For example, in figure 7 of FC, even for the SCDM model, the expected number of \ion{O}{8} absorption lines with $N^{i} > 10^{15}\ cm^{-2}$ and $z < 0.3$ is equal to or below one.

This method illustrates the possibility of constraining the cosmological parameters by observing more X-ray bright quasars with longer exposure time. For example, if a quasar is brighter than S5 0836+710 by a factor of two, by increasing the exposure time to $100$ ksec, the minimum detectable column density for \ion{O}{8} can be as low as $N^{i}_{0} = 3 \times 10^{16}\ cm^{-2}$. For a low density cold dark matter model (LCDM) with $\Omega_{0} = 0.3$ and $\Omega_{\Lambda} = 0.7$ \footnote{Here $\Omega_{\Lambda} \equiv \Lambda/H_{0}^{2}$ where $\Lambda$ is the cosmological constant}, the expected number of \ion{O}{8} absorption lines is $0.42$, given a reasonable cluster gas fraction ($\sim 0.2$) and metallicity ($\sim 0.5 Z_{\odot}$ for oxygen). The Poisson statistic shows that the probability of observing at least one line is $95\%$ if the expectation is three. This means by observing $\sim 7$ sources which are twice brighter than S5 0836+710, we should have $95\%$ chance of detecting at least one \ion{O}{8} absorption line with an exposure time of $100$ ksec. Previous observations show that at low and moderate redshifts, there are several tens of quasars having a flux higher than $5 \times 10^{-11}\ ergs\ cm^{-2}s^{-1}$ between $2 - 10$ keV, or $2 \times 10^{-11}\ ergs\ cm^{-2}s^{-1}$ between $0.1 - 2.4$ keV (see, e.g., \citealp{bys97,ybs98,rtu00}). By increasing the number observations of bright quasars at low and moderate redshifts, there is a significant chance of detecting oxygen absorption lines. 

Similar constraints can be made with current and future missions, such as {\sl XMM} and {\sl Constellation-X}. For example, {\sl XMM} and {\sl Constellation-X} are able to detect \ion{O}{8} at a column density as low as $7.9 \times 10^{15}\ cm^{-2}$ and $10^{15}\ cm^{-2}$, respectively (see FC). Given such a low column density, we will detect as many as eight or ten \ion{O}{8} absorption lines for a moderately luminous quasar at redshift of $z \sim 1$, respectively. Eventually this method may help us to understand the distribution of the baryonic matter in the highly overdense regions.

\subsection{X-ray Gunn-Peterson Test \label{xgpt}}

Recently large-scale cosmological hydrodynamic simulations have shown that a large amount of the hot/warm IGM ($70\%\sim80\%$) lies in the regions at overdensity $5 < \delta < 200$ and temperature $10^{5} < T < 10^{7}$ (see, e.g., \citealp{cos99,dco00}). Here $\delta = \rho/\left<\rho\right> - 1$ and $\left<\rho\right>$ is the mean matter density. These regions are typically the extended filamentary structures connecting regions of high mass concentration. The X-ray Gunn-Peterson test may be applied to examine the possible ``absorption trough'' in the X-ray continuum spectrum of a distant quasar, produced by these diffuse structures. \citet{sba80} and \citet{aem94} discussed the absorption produced by a uniformly distributed IGM. \citet{mar99} proposed that the X-ray Gunn-Peterson test can be used on a particular giant filament along the line of sight, in the direction of Aquarius. Based on numerical simulations and observations of the Ly$\alpha$ forest systems we know that instead of a uniform distribution, these filaments are actually clumpy structures in the universe. However, to simplify, we assume a uniformly distributed IGM and constrain its properties following \citet{sba80} and \citet{aem94}. 

Consider the X-ray spectrum of a background quasar located at redshift $z_{0}$, the frequency of a photon emitted at $\nu_{0}$ will be $\nu = \nu_{0} (1+z)^{-1}$ when the photon reaches the Earth. The optical depth at frequency $\nu$ (observer frame) is \citep{spi78}
\begin{equation}
\label{eq:tau}
\tau(\nu) = \int_{0}^{z_{0}} n_{i}\left(z\right)\sigma_{i}\left(\nu^{\prime}\right)\frac{dl}{dz}dz
\end{equation} where $n_{i}(z)$ is the comoving number density of absorbing ion $X^{i}$ at a redshift of $z$ and $\sigma_{i}$ is the ion absorption cross section at frequency $\nu^{\prime} = \nu\left(1+z\right)$. The path length is $dl$.

There are three dominant absorption processes contributing to the optical depth in equation (\ref{eq:tau}): resonant absorption due to the atomic transitions, continuum absorption (edge) due to the bound-free transitions, and electron scattering. Electron scattering is independent of energy in the X-ray band and the only effect is uniformly lowering the X-ray continuum without altering the spectral shape. We ignore this effect because the absorption produced by this process will not be detectable without knowledge of the intrinsic quasar spectrum. 

To simplify we assume that the IGM is uniformly distributed hot gas in collisional equilibrium, at temperature $T \ge 10^{6}$ K. In such an environment, the hydrogen- and helium-like O, Ne, Mg, Si, and \ion{Fe}{17}-\ion{Fe}{26} are the most abundant ions (for ionization fractions in collisional equilibrium, see \citealp{mmc98}). For the continuum absorption we include all the K-edges (i.e., 18 K-edges) and L-edges for all the ions with L-shell electrons (i.e., 13 L-edges). We obtain the threshold energies and cross sections from \citet{vya95}. For resonant absorption, we select those transitions with oscillator strength greater than $0.1$. This results in a total of $65$ lines. The energy level and oscillator strength for each resonance line are from \citet{vvf96}. The total optical depth $\tau_{\nu}$ then can be calculated based on equation (\ref{eq:tau}).

The quasar spectrum $f_{\nu}$ can be described as \citep{aem94}
\begin{equation}
\label{eq:fit}
f_{\nu} = Af\left(N_{H}\right)\left(\frac{E}{1\ keV}\right)^{-\Gamma}\exp\left[-\tau\left(\Omega_{b}, T\right)\right]
\end{equation} Here $A$ is a normalization constant and $f\left(N_{H}\right)$ is the the gas absorption where $N_{H}$ is the hydrogen column density. The exponential term is the absorption of the IGM at a baryon fraction $\Omega_{b}$ and temperature $T$. We first fit the observed spectrum with $\Omega_{b}=0$. Then for each $\Omega_{b}$ and $T$ we fit the observed spectrum with equation (\ref{eq:fit}) by varying $A$, $N_{H}$ and $\Gamma$. We then calculate the statistic 
\begin{equation}
\Delta\chi^{2}(\Omega_{b}, T) =  \chi^{2}(\Omega_{b}, T) -  \chi^{2}(\Omega_{b}=0)
\end{equation}  The $\Delta\chi^{2}(\Omega_{b}, T)$ should follow the {\it chi}-square distribution with two degrees of freedom. In this way we can set confidence levels on $\Omega_{b}$ and $T$.

We select the MEG spectrum of S5 0836+710 because the MEG covers the entire range from oxygen to iron. Figure~\ref{f6}a displays the $3\sigma$ contour of $\Delta\chi^{2}(\Omega_{b}, T)$ for a flat, low-density cosmological model (LCDM, $\Omega_{0} = 0.3, \Omega_{\Lambda} = 0.7$). The thick lines indict an abundance ($Z_{\odot}=0.5$) and the thin lines are for a low-abundance IGM ($Z_{\odot}=0.1$). For comparison we also show the constraint from the distortion of the cosmic microwave background (CMB) spectrum (which constrains the Compton $y$-parameter, the dotted line) as measured by the {\sl COBE} satellite (\citealp{mcc94,fcg96}). The regions to the left of the solid and dotted lines are allowed regions. 

\placefigure{f6}

We also display the total baryon density deduced from standard big bang nucleosynthesis (BBN) and the cosmic microwave background (CMB). BBN implies $\Omega_{b}h^{2}_{70} = 0.0388\pm0.0037$ ($95\%$ confidence level, hereafter cl) (see, e.g., \citealp{bnt00}). The recent measurements of CMB anisotropy by {\sl BOOMERANG} and {\sl MAXIMA} gave $\Omega_{b}h^{2}_{70} = 0.065^{+0.018}_{-0.016}$ ($95\%$ cl) \citep{jab00}. We notice that while the $\Omega_{b}$ measured in this way reflects the small angle (subdegree) anisotropies, the constraints based on the the Compton $y$-parameter reflects the CMB anisotropies at larger scales. We display both BBN and CMB results and the shadowed area shows the $95\%$ cl region. 

From the figure we can find that while for a low-abundance IGM the X-ray Gunn-Peterson test is not able to put strong constraints on the properties of the IGM, for a uniformly enriched IGM with $Z_{\odot}=0.5$, marginal constraints can be made by combining the the Gunn-Peterson test and the Compton $y$-parameter from the CMB anisotropies. For example, for the CMB predicted baryon density, the temperatures of the IGM are constrained in a range with $10^{6.1} \lesssim T \lesssim 10^{7.7}$ K: temperatures above this range are ruled out by the CMB spectrum, and at a temperature below this range the IGM will produce strong distortion in the quasar spectrum. In figure~\ref{f6}b we illustrate such a distortion in the MEG spectrum of S5 0836+710, based on a LCDM model. The dotted line represents the absorbed spectrum by an IGM with $Z_{\odot}=0.5$ at $T=10^6$ K and $\Omega_{b}h^{2}_{70} = 0.08$. Given such an IGM, the model severely deviates from the observed spectrum. 

Based on the {\sl ROSAT} PSPC observations of three high-redshift quasars, \citet{aem94} concluded that the X-ray opacity of a hot diffuse IGM is too small to constrain the IGM. For instance, they found that their X-ray Gunn-Peterson test did not constrain the IGM temperature above $10^{4}$ K with $\Omega_{b} = 0.06$ and solar abundances (Figure 11 in their paper). Our analysis indicates that the constraint can be improved significantly even with half-solar abundances, due to the high energy resolving power of {\sl Chandra} HETGS. Our results are similar to their predictions with {\sl Chandra} Low-Energy Transmission Grating Spectrometer (LETG), based on a simulated observation of 3C 273. However, due to the limitation of the small effective area, we are not able to constrain the IGM with low metallicity based on current data. In future, {\sl Constellation-X} will increase the effective area by two magnitudes over {\sl Chandra} HETGS and may eventually reveal the properties of the IGM.

\section{Summary}

In this paper we report the {\sl Chandra} observations of two high-redshift, radio-loud quasars, S5 0836+710 and PKS 2149-306.

\begin{enumerate}

\item The photon index ($\Gamma$) for S5 0836+710 is 1.39, consistent with previous observations with {\sl ROSAT} and {\sl ASCA}. Compared with previous observations, PKS 2149-306 has a rather soft spectrum with $\Gamma=1.255$. Both photon indexes are consistent with the fact that RLQs have flatter spectra compared with RQQs. We also find excess continuum absorption (above the Galactic value) in S5 0836+710. The flux of S5 0836+710 is higher than in previous {\sl ASCA} observations and puts this source as one of the most luminous objects at high redshift.

\item No significant absorption or emission feature is found in both sources at or above $\pm3\sigma$ level. We put constraints on the possible emitters or absorbers. For example, Fe K emission lines were found in several RLQs \citep{rtu00}, although not as frequently as in RQQs. We put tight upper limits on EW, as low as $\sim 10\ eV$ in both sources. We do not find the emission feature reported by Y99 in the spectrum of PKS 2149-306 around $5\ keV$. We also examine the possible intervening systems and give upper limits on the absorbers' ion column density. 

\item We discuss the possibility of constraining the properties of the IGM. We develop a method to constrain cosmological parameters based on the X-ray forest method. However, we are not able to set any meaningful constraints based on current data. The X-ray Gunn-Peterson test was also performed and we find that, by combining with the constraints from the CMB Compton $y$-parameter, this method can be applied to constrain a uniform, enriched IGM.

\end{enumerate}

\acknowledgments{TF thanks the support from MIT/CXC team, especially David Davis, Dan Dewey, John Davis, John Houck and David Huenemoerder. This work is supported in part by contracts NAS 8-38249 and SAO SV1-61010. Support for GLB was provided by NASA through Hubble Fellowship grant HF-01104.01-98A from the Space Telescope Science Institute, which is operated by the Association of Universities for Research in Astronomy, Inc., under NASA contract NAS 6-26555.}

\clearpage

\figcaption[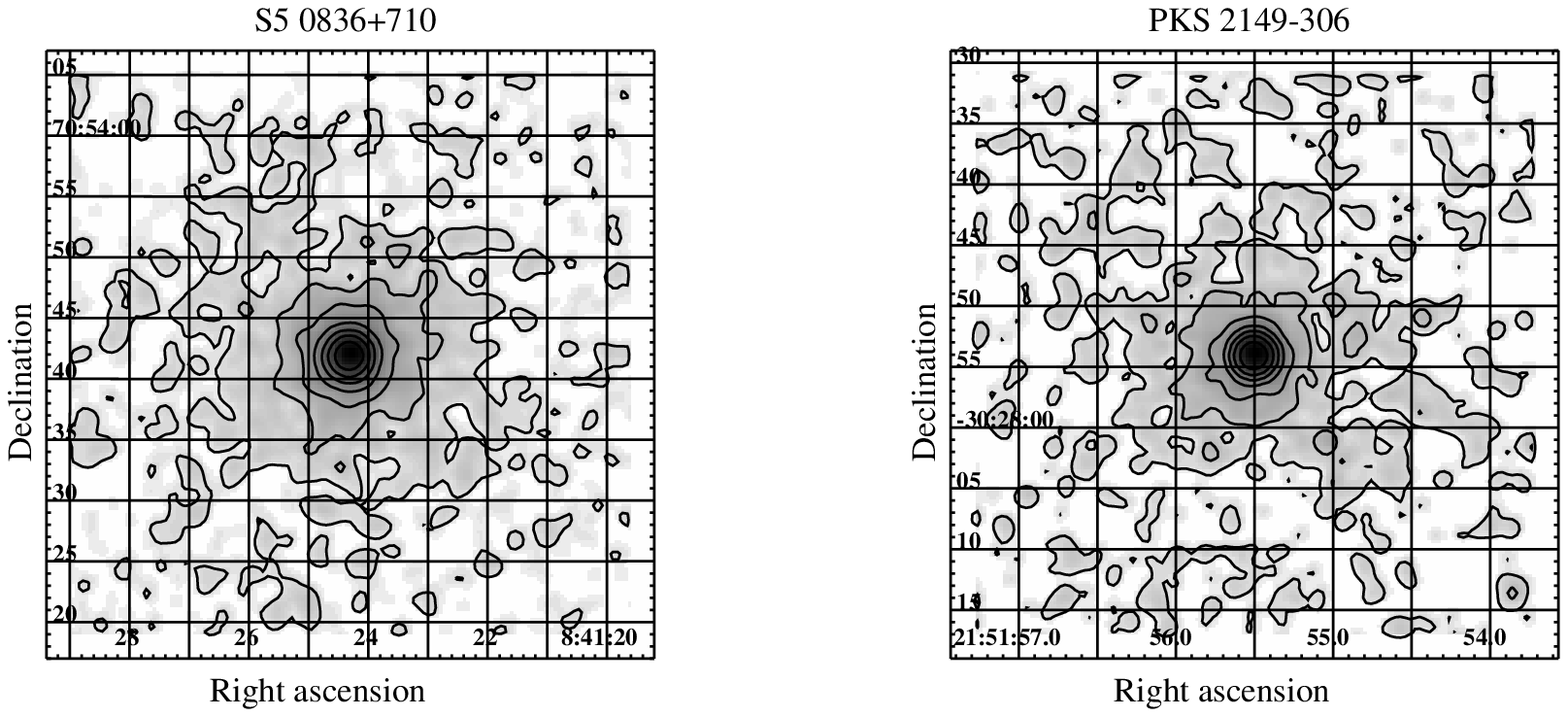]{Isophotal contour map for the zeroth-order images of S5 0836+710 (left) and PKS 2149-306 (right).The region size is $50^{\prime\prime} \times 50^{\prime\prime}$. The images are smoothed with a Gaussian ($\sigma=1.2^{\prime\prime}$) for visual clarity. Both images can be well fitted by a single Gaussian with $\sigma \approx 0.38^{\prime\prime}$ for S5 0836+710 and $\sigma \approx 0.35^{\prime\prime}$ for PKS 2149-306. \label{f1}}

\figcaption[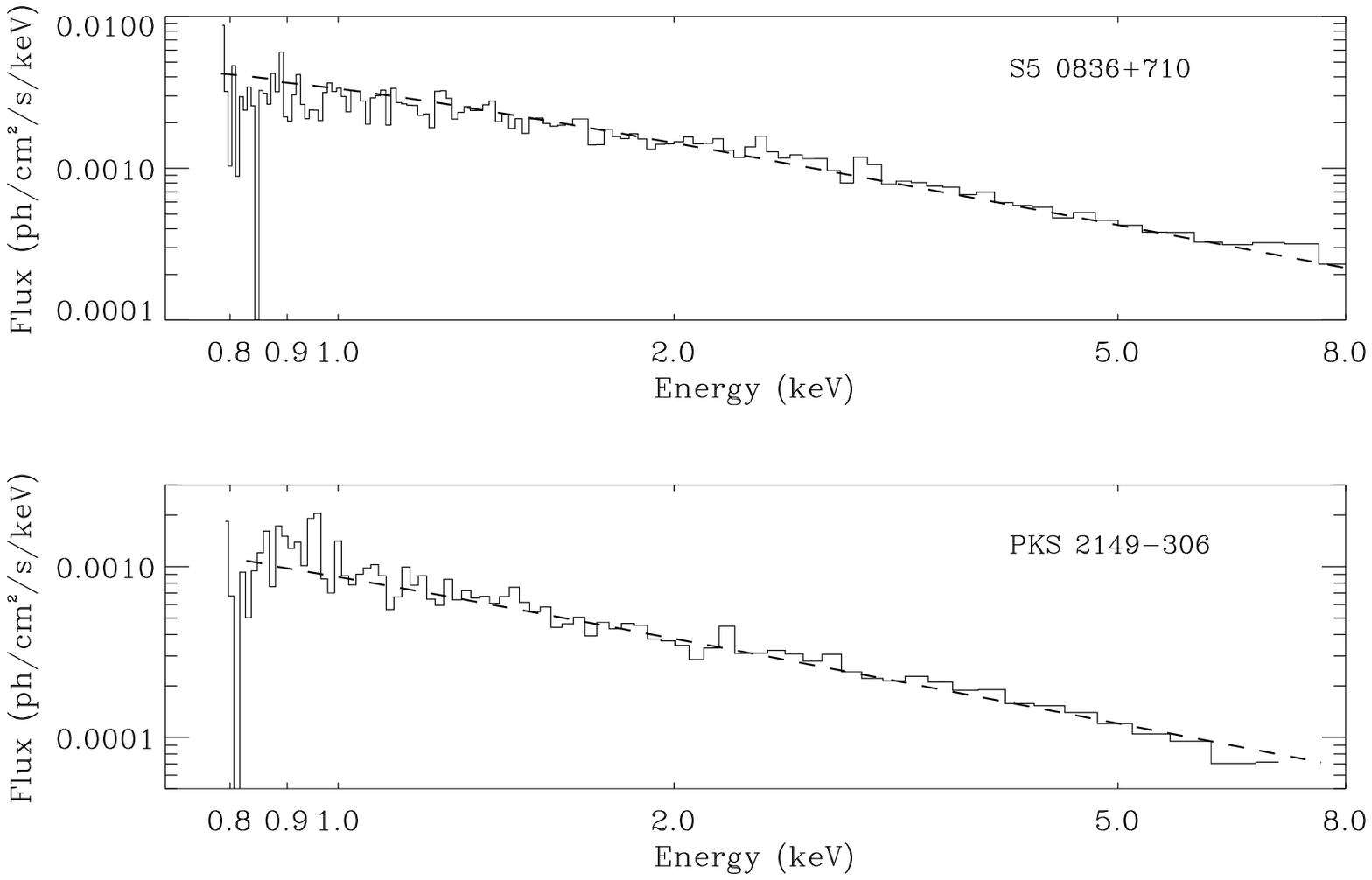]{HEG spectra of S5 0836+710 (top panel) and PKS 2149-306 (bottom panel). The solid line and dotted line in each panel are the observed and fitted spectrum, respectively. The spectral model is an absorbed power-law and the best-fit parameters are from Table~\ref{t2}. \label{f2}}

\figcaption[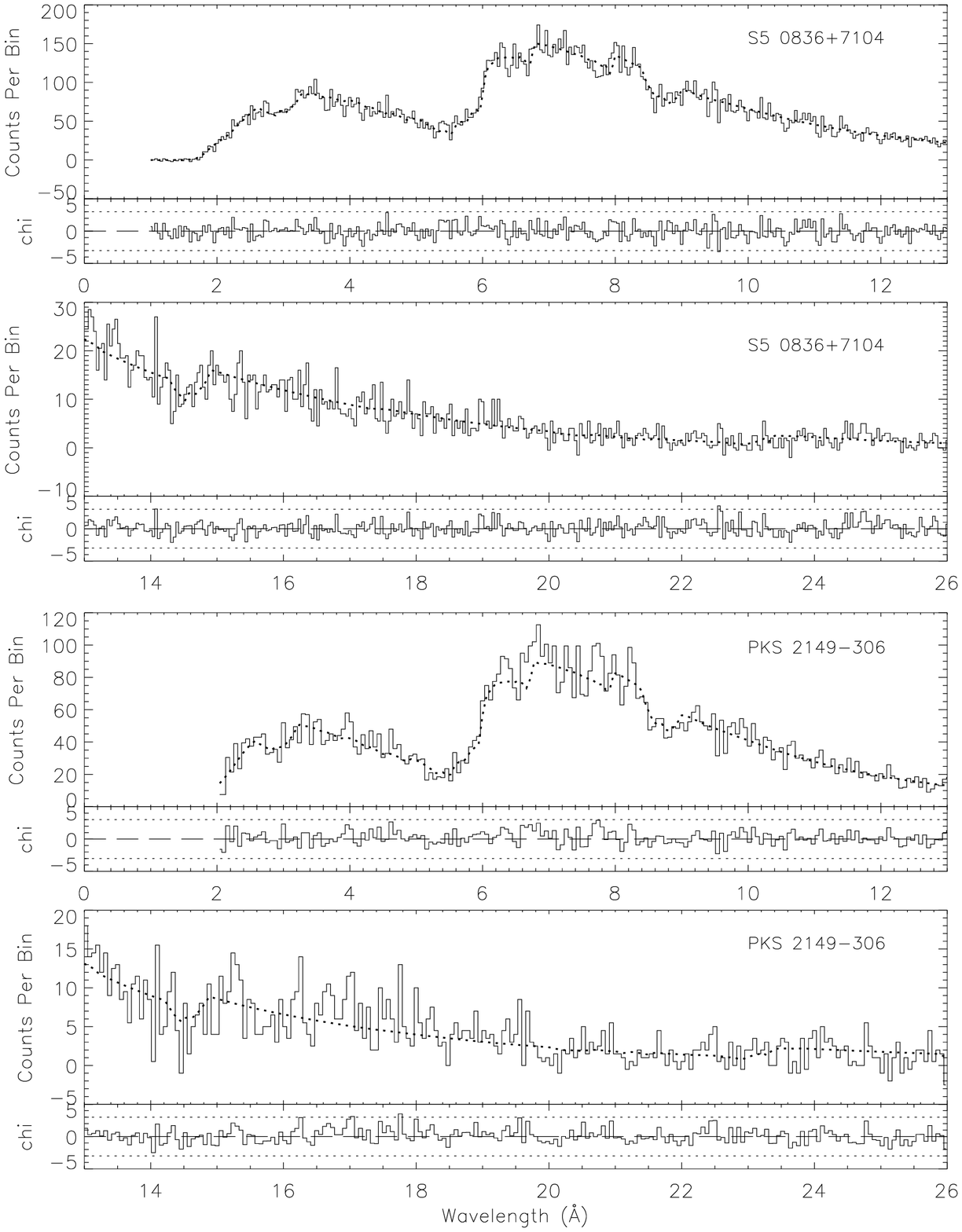]{MEG spectra of S5 0836+710 (top two panels) and PKS 2149-306 (bottom two panels) in wavelength space. The bin size is $0.04\AA$ for S5 0836+710 and $0.06\AA$ for PKS 2149-306. The observed spectra are first fitted with an absorbed power-law (same parameters as in HEG spectra), the residual is then fitted with a 5-order polynomial. The bottom figure in each panel shows the $\chi$, and the two dotted lines indicate the $\pm3\sigma$ level. \label{f3}}

\figcaption[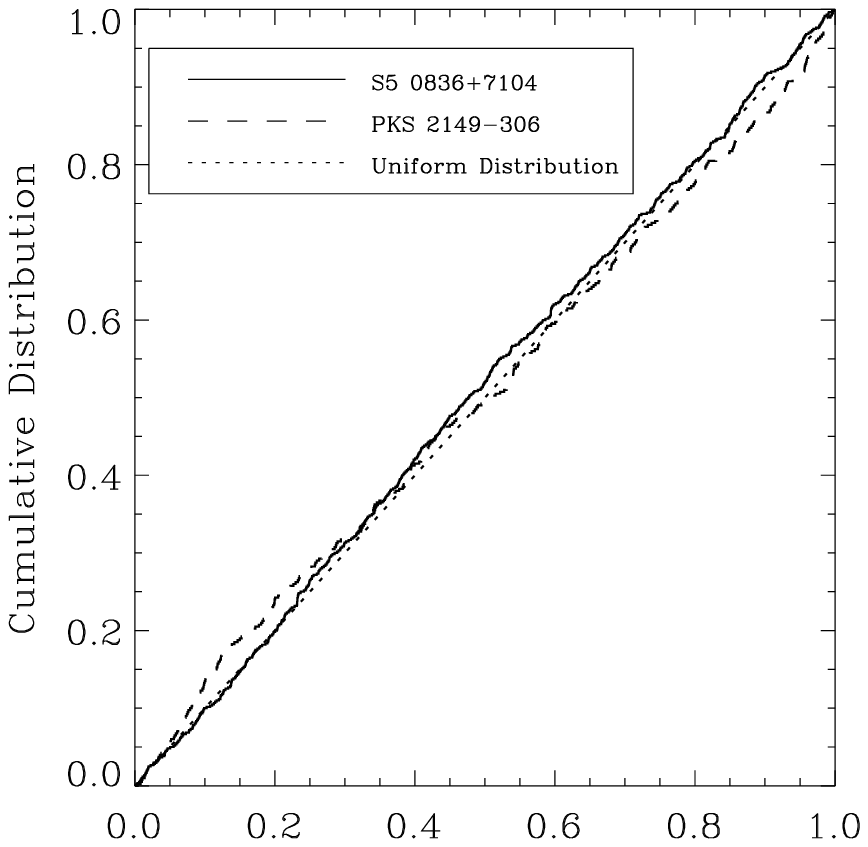]{The cumulative distribution of the Poisson Probabilities fpr S5 0836+710 (solid line), PKS 2149-306 (dashed line) and the uniform distribution (dotted line). The KS test shows that the null hypothesis that they are the same distributions is accepted at a significant level of $0.59$ for S5 0836+710 and $0.32$ for PKS 2149-306.\label{f4}}


\figcaption[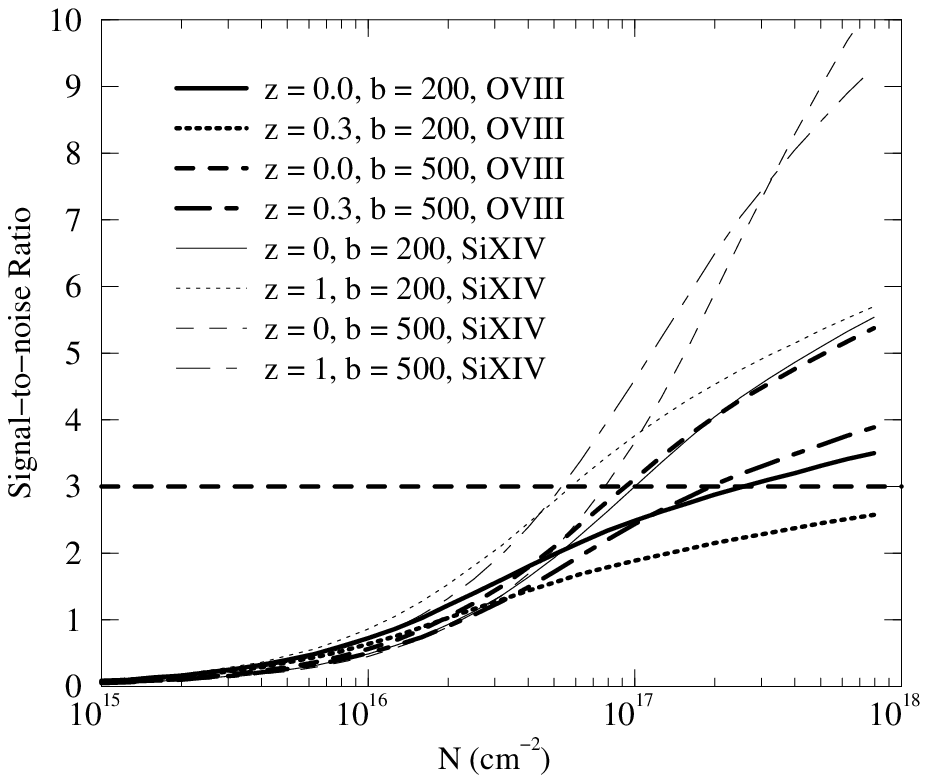]{Given the spectrum of S5 0836+710, we calculate the signal-to-noise ratio of \ion{Si}{14} and \ion{O}{8} absorption lines with different redshift ($z$) and velocity dispersion ($b$, in units of $km\ s^{-1}$). \ion{Si}{14} is based on HEG spectrum and \ion{O}{8} is based on MEG spectrum. Since MEG has no effective area below $0.5\ keV$, the maximum redshift of O VIII is $0.3$. The horizontal line is the $3\sigma$ level. To obtain a Si XIV or O VIII absorption line in the spectrum of S5 0836+710 at $3\sigma$ level, the ion column density must be at least $5 \times 10^{16}\ cm^{-2}$ and $8 \times 10^{16}\ cm^{-2}$, respectively. \label{f5}}

\figcaption[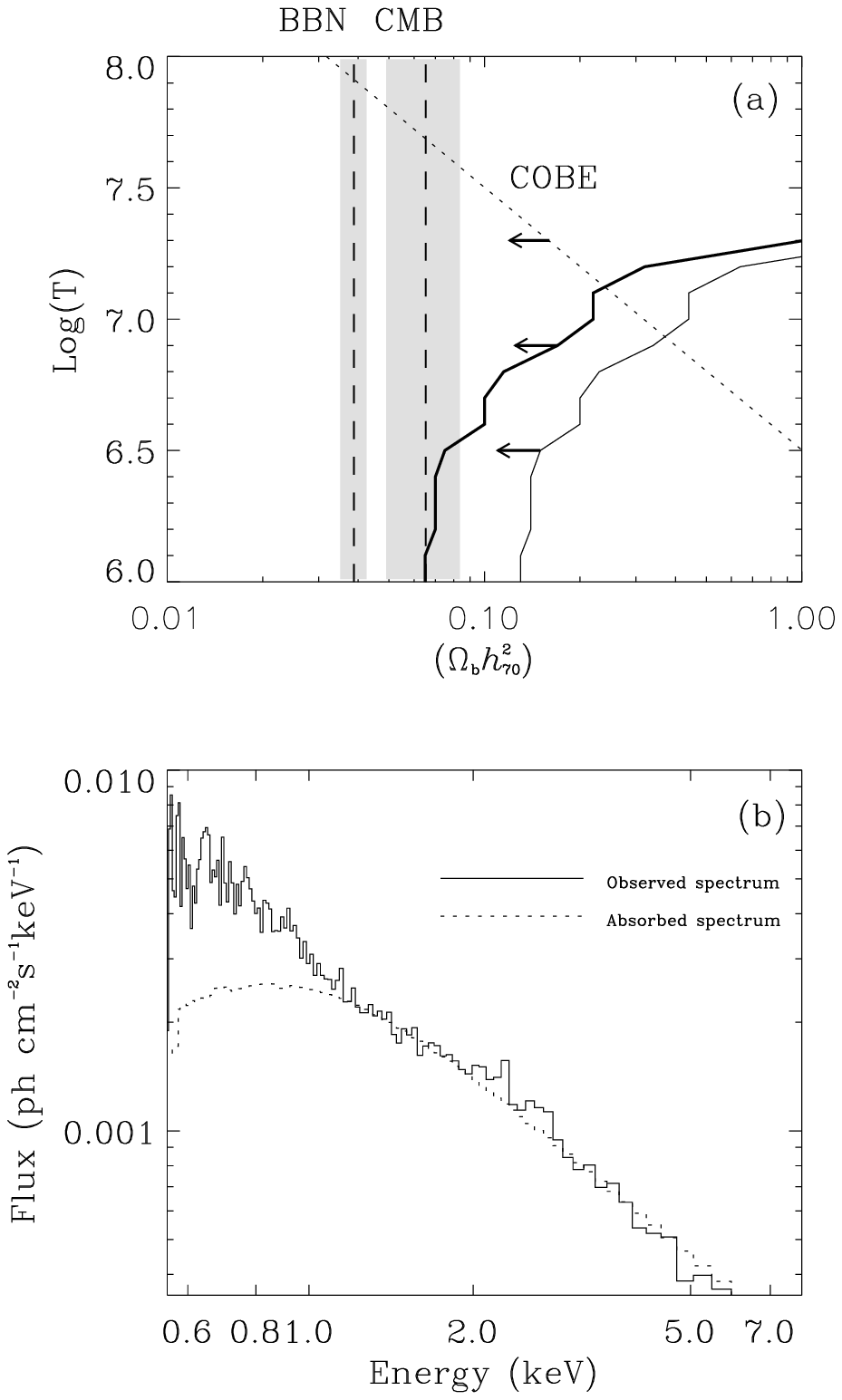]{(a) Constraints on $\Omega_{b}$ and the IGM temperature from the X-ray Gunn-Peterson test, based on the MEG spectrum of S5 0836+710. The top panel shows the $3\sigma$ contour of ($\Omega_{b}h_{70}^{2}$) for each IGM temperature. The thick line indicts an enriched IGM with $Z_{\odot}=0.5$ and the thin line is for a low-abundance IGM ($Z_{\odot}=0.1$). The dotted line shows the constraint form the distortion of the CMB spectrum (the Compton $y$-parameter) as measured by the {\sl COBE}. The regions to the left of the solid (LCDM) and dotted ({\sl COBE}) lines are acceptable (indicated by arrows). (b) The solid line is the observed MEG spectrum of S5 0836+710. The dotted line is the distorted spectrum absorbed by an enriched IGM ($Z_{\odot}=0.5$) with a temperature of $10^{6}$ K and $\Omega_{b}h_{70}^{2} = 0.08$. \label{f6}}

\clearpage


\begin{deluxetable}{cccccc}
\tablecaption{{\sl Chandra} Observation Log
\label{t1}}
\tablehead{ \colhead{Observation ID} & \colhead{Object} & 
\colhead{Redshift} & \colhead{$N_{H gal}$\tablenotemark{a}} &
\colhead{Start Date} & \colhead{Exposure Time ($ksec$)}}
\startdata 
01450 & S5 0836+710 & 2.170 & 2.83 & 10/17/99 & 58.7 \\
00336 & PKS 2149-306 & 2.345 & 1.91 & 11/18/99 & 37.3 \\
01481 & PKS 2149-306 & 2.345 & 1.91 & 11/20/99 & 55.5 \\
\enddata
\tablenotetext{a}{Galactic absorption from \citet{mll96} (S5 0836+7104) and \citet{dlo90} (PKS 2149-306), in units of $10^{20}\ cm^{-2}$.}
\end{deluxetable}

\begin{deluxetable}{lccccrc}
\tabletypesize{\footnotesize} 
\tablewidth{0pt}
\tablecaption{{\sl Chandra} Spectral Fit
\label{t2}}
\tablehead{\colhead{Object} & 
\colhead{$N_{H}$ ($10^{20}\ cm^{-2}$)} & \colhead{Photon Index ($\Gamma$)} &
\colhead{$A_{pl}$\tablenotemark{a}} & \colhead{$\chi^{2}_{red}$/dof} & 
\colhead{$f_{2-10}$\tablenotemark{b}} & \colhead{$\log L$\tablenotemark{c}} \\ \colhead{} & \colhead{} & \colhead{($0.5 - 8\ keV$)} &  \colhead{} &  \colhead{} & \colhead{} & \colhead{}}
\startdata 
S5 0836+710 & $7.0\pm1.2$\tablenotemark{d} & $1.388\pm0.012$ & $3.96^{+0.04}_{-0.07}$ & 1.072/605 & 26.42 & 47.43$h_{70}^{-2}$\\
PKS 2149-306 & 1.9  & $1.255\pm0.020$ & $0.910^{+0.015}_{-0.011}$ & 0.965/605 & 7.58 & 46.94$h_{70}^{-2}$\\
\enddata
\tablenotetext{a}{Normalization at $1\ keV$ (observer's frame) in units of $10^{-3}\ photons\ cm^{-2}s^{-1}\ keV^{-1}$}
\tablenotetext{b}{Absorbed flux between $2$ and $10 \ keV$ (observer's frame) in units of $10^{-12}\ ergs\ cm^{-2}s^{-1}$}
\tablenotetext{c}{Intrinsic luminosity between $2$ and $10 \ keV$ (quasar frame) in units of $ergs\ s^{-1}$, $q_{0} = 0.5$}
\tablenotetext{d}{All errors are quoted at $90\%$ confidence}
\end{deluxetable}

\clearpage
\begin{deluxetable}{crrrcrr}
\tablecolumns{7}
\tablewidth{0pc}
\tablecaption{The Maximum Equivalent Width of Possible Lines \tablenotemark{a} \label{t3}}
\tablehead{\colhead{} & \colhead{} & \multicolumn{2}{c}{HEG} & \colhead{} & \multicolumn{2}{c}{MEG} \\ \cline{3-4} \cline{6-7} \colhead{Object} & \colhead{Energy\tablenotemark{b}} & \colhead{$\sigma = 0$\tablenotemark{c}} & \colhead{$\sigma = 0.1$} & \colhead{} & \colhead{$\sigma = 0$} & \colhead{$\sigma = 0.1$}}
\startdata
S5 0836+710 & 6.4 & 9 & 72 &   & 8 & 49 \\
             & 6.9 & 14 & 100 &   & 15 & 84 \\
\cline{1-7}
PKS 2149-306 & 6.4 & 10 & 83 &   & 8 & 53 \\
             & 6.9 & 12 & 94 &   & 12 & 69 \\
	     & 17.0 & 67 & 188 &   & 97 & 215
\enddata
\tablenotetext{a}{The equivalent width is measured in the rest frame of quasar ($eV$)} \tablenotetext{b}{Line energy, in quasar frame ($keV$). The Fe K emission lines are located between $6.4\ keV$ and $6.9\ keV$, the $17.0\ keV$ line is the redshifted Fe-K line reported by Y99.} \tablenotetext{c}{Intrinsic line width ($keV$)}
\end{deluxetable}

\clearpage
\begin{figure}
\plotone{f1.eps}
\end{figure}

\clearpage
\begin{figure}
\plotone{f2.eps}
\end{figure}

\clearpage
\begin{figure}
\plotone{f3.eps}
\end{figure}

\clearpage
\begin{figure}
\plotone{f4.eps}
\end{figure}

\clearpage
\begin{figure}
\plotone{f5.eps}
\end{figure}

\clearpage
\begin{figure}
\epsscale{0.8}
\plotone{f6.eps}
\end{figure}

\end{document}